# Degradation Dynamics of Perovskite Solar Cells Under Fixed Reverse Current Injection

*Fangyuan Jiang[1], Haruka Koizumi[1], Hannah Contreras[1], Rajiv Giridharagopal[1], Akash Dasgupta[1], Zixu Huang[1], Ryan A. DeCrescent[2,3], Kell Fremouw[3,4], Michael D. McGehee[2,3], Neal R. Armstrong[5], David S. Ginger[1]\**

[1]Department of Chemistry, University of Washington, Seattle, WA 98195, USA
[2]Chemical and Biological Engineering, University of Colorado Boulder, Boulder, CO 80309, USA
[3]Renewable and Sustainable Energy Institute (RASEI), University of Colorado Boulder, Boulder, CO 80303, USA
[4]Materials Science and Engineering, University of Colorado, Boulder, CO 80303, United States
[5]Department of Chemistry and Biochemistry, University of Arizona, Tucson, AZ 85721, USA

\* Corresponding author email: dginger@uw.edu

**Abstract**

Previous studies of reverse-bias stability in perovskite solar cells have focused primarily on voltage-controlled reverse-bias tests. Here we instead present an investigation of perovskite solar cell degradation under well-defined, constant reverse-current stress. We show that the choice of hole-transport layer dictates the dominant degradation pathway: cells using thick poly(triphenylamine) (PTAA) layers with better indium-doped tin oxide (ITO) coverage can tolerate high reverse bias but quickly undergo catastrophic breakdown under fixed reverse current near their one-sun maximum power-point. In contrast, cells modified with the phosphonic-acid interface layer MeO-2PACz, with poorer ITO coverage compared to PTAA, exhibit soft, gradual, and largely recoverable degradation, regardless of the shading conditions. For MeO-2PACz devices, degradation increases with both current magnitude and duration. Importantly, when normalized by injected charge (current times duration), lower currents applied over longer times cause more severe degradation than higher currents over shorter periods. Combining electrical measurements with spatially resolved photoluminescence imaging, we argue against shunt formation and instead support an ion- and charge-mediated interfacial electrochemical degradation mode.

**TOC**

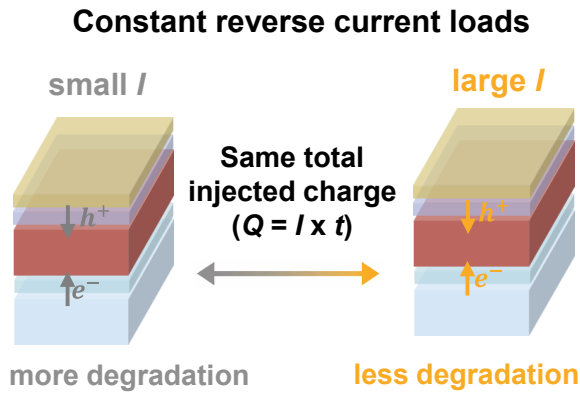

**Constant reverse current loads**

small *I*      large *I*

Same total injected charge ($Q = I \times t$)

more degradation      less degradation

**Introduction**

The operational instability of perovskite solar cells under realistic operating conditions remains a key bottleneck for their commercial deployment.[1,2] One critical yet underexplored stressor is current injection under reverse bias, which arises when a solar cell within a series-connected module generates less photocurrent than its neighbors, possibly due to non-uniform (partial) shading. The remaining illuminated cells operating near their maximum power point force the module operating current (typically around the current density at maximum power point $J_{mpp}$) through the shaded cell in the "wrong" direction.[3–5]

Reverse-bias instability poses a challenge not only for perovskite photovoltaics but also for established PV technologies such as silicon (Si) and cadmium telluride (CdTe).[6,7] A range of device and materials engineering strategies have been developed to mitigate its impact, including the bypass diode strategy widely implemented in commercial Si PV modules,[6,8–11] the long-strip cell layouts adopted in First Solar's thin-film CdTe modules,[12] and the interdigitated back contact (IBC) architecture seen in SunPower's Si PV technology.[13,14] For perovskite solar cells in particular, notable progress has been made to improve the breakdown voltage ($V_{rb}$) through interfacial and material engineering,[15–20] as well as through device design such as two-terminal perovskite – Si tandem configurations[21–24] and mesoscopic architectures employing carbon electrodes.[25] From this standpoint, perovskite solar cells now appear to have viable pathways toward module integration in layouts where bypass diodes can be incorporated.

However, to date there has been little if any work exploring perovskite solar cells stability under fixed reverse current. Such studies may inform what factors contribute to degradation, while at the same time providing insight into the technological directions that might enable perovskite cells to sustain soft, spatially uniform reverse-current flow without the formation of localized hotspots or irreversible damage. If feasible, such behavior would in theory allow module operation with fewer, or even without, bypass diodes, which could be important to many large-area thin-film module manufacturing approaches. Although a growing body of literature has examined perovskite stability under fixed reverse bias *or* quick $J – V$ scans,[16–18,25–29] few studies have focused on their degradation under fixed reverse current stress conditions.[30] From previous studies, it is clear that single-junction perovskite devices degrade under fixed reverse biases, even at current densities below ~1 mA/cm$^2$,[16,26] far lower than $J_{mpp}$ (typically between 15 – 30 mA/cm$^2$) that an unprotected cell would need to tolerate for hours under real-world operating conditions.

Here, we directly study perovskite solar cell degradation under well-defined, fixed reverse-current injection condition. We compare cells with a thick (~35 nm) PTAA hole-transport layer (HTL) covering the indium-doped tin oxide (ITO) bottom contact, to those where the ITO contact is modified by the popular methoxy-substitute carbazole-phosphonic-acid interface modifier MeO-2PACz, which is known to lead to heterogeneous coverage.[31,32] Although devices with thick PTAA layers can sustain higher *reverse bias*, they can break down catastrophically at $J_{mpp}$-level reverse currents. In contrast, MeO-2PACz-based cells, which feature a lower electron injection barrier and do not appear to fully block the contact with the underlying ITO contact,[16] exhibit gradual and largely recoverable degradation that scales to first order with the cumulative injected charge (current density times duration). The degradation dynamics depend on both the magnitude and duration of the injected current. A closer inspection reveals an important result: we find that lower currents applied over longer durations result in more severe degradation than higher currents for shorter durations for equal total injected charge. Finally, we show that progressive photoluminescence (PL) quenching accompanies device degradation, with increased charge injection corresponding to greater PL quenching. Together, these observations are consistent with previously proposed models[33,34] in which the current flowing through a p-i-n perovskite diode consists of two parallel pathways: (i) charge transport through electronic bands (normal forward or reverse bias diode operation); (ii) ion motion coupled with electrochemical processes occurring at the top and bottom contacts that are not fully passivated. Under high reverse currents for shorter durations, these electrochemical processes may be transport limited and constitute a smaller fraction of the total currents, causing less degradation compared with lower currents applied for longer durations.

**Results and Discussions**

We begin by investigating the stability of a typical *p-i-n* structured perovskite solar cell under constant reverse-current injection, using a current density ~equal to the maximum power point ($J_{mpp}$ = 19 mA/cm$^2$). **Figure 1a** shows our cell architecture: the absorbing layer is a ~460-nm thick perovskite with the composition Cs$_{0.22}$FA$_{0.78}$Pb(I$_{0.85}$Br$_{0.15}$)$_3$ with 3 mol% MAPbCl$_3$ added to the solution (hereafter referred to as Cs22Br15Cl).[35] FA and MA denote formamidinium and methylammonium, respectively. For the HTL, we employ a 35-nm-thick PTAA layer which has previously been shown to effectively planarize the transparent conductive oxide substrate and suppress electron injection under reverse bias.[16] Both properties likely contribute to devices survival under high reverse *bias* stress,[16] as would be suitable for a module architecture employing bypass diodes. We use C$_{60}$/BCP as the electron transport layer and Au as the rear electrode.[16,36] During reverse-bias measurements, we ground the Au electrode and apply negative bias to the ITO.

Consistent with our previous results, the PTAA cells exhibit high $V_{rb}$ (> |-15| V, **Figure S1**).[16] These results confirm the absence of significant pinholes or fabrication defects. However, sourcing $J_{mpp}$ results in nearly immediate, catastrophic failure, accompanied by burn marks on the cells that are visible by naked eye (**Figure 1b**). Sourcing $J_{mpp}$ through a single cell is a markedly different condition from doing a reverse bias scan: it allows sourcing of whatever (possibly very large) voltage is required to drive the reverse current,

which can easily exceed 100 V (over 200 V in the case of CdTe PV module[37]) in an unprotected series-connected module. The burn marks resulting from this measurement are consistent with reverse bias damage reported across many solar cell technologies where local hot spots lead to rapid heating and cascading material failure.[27,28,38–41] Using scanning electron microscopy, we find microscale volcano-type features, with the Au electrode layer peeling off from affected regions and locally damaged perovskite (**Figure 1c** and **Figure S2**).

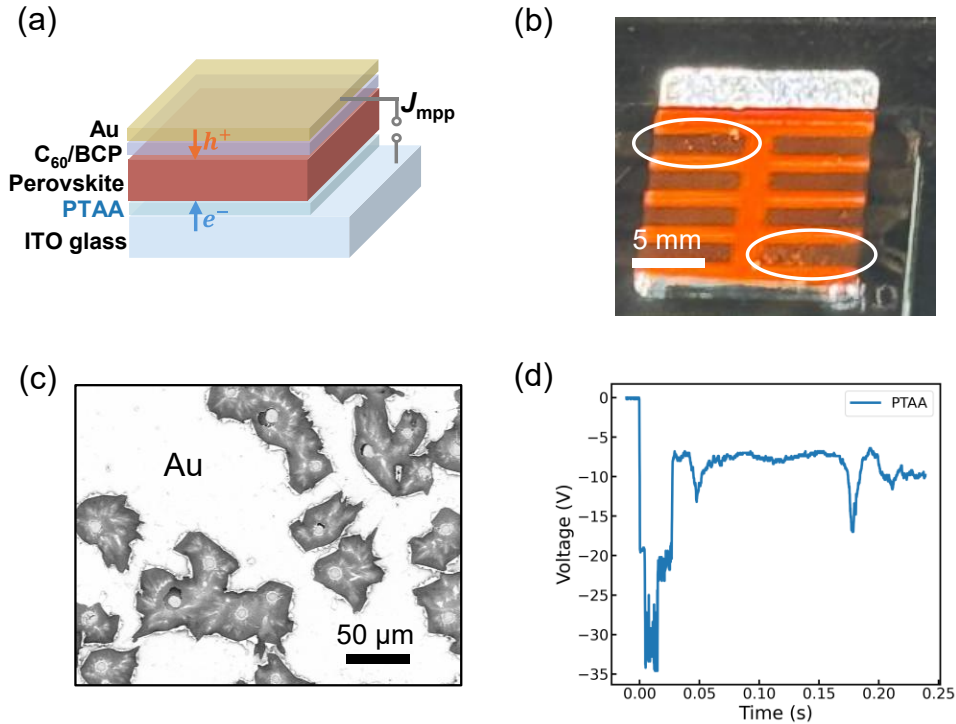

**Figure 1.** (a) Device structure of perovskite solar cell using ~35 nm – thick PTAA hole transporting layer; (b) Optical image showing the burn-out marks in PTAA cells after current injection stressing; (c) Scanning electron microscopy showing the micro-scale "volcano" features at these burn-out regions; (d) The evolution of reverse bias over time when the cell is sourced with $J_{mpp}$ (19 mA/cm$^2$) in the dark, as recorded by oscilloscope.

To better understand this process, we use an oscilloscope to monitor the bias sourced across the cell with about 0.1 ms time resolution immediately after applying the fixed reverse current. Within the first 50 ms, the voltage across the cell rapidly rises to nearly -35 V, followed by a rapid drop to -7 V (**Figure 1d**). In perovskite solar cells that have a sufficient amount of contact area between the perovskite and the ITO electrode, electrochemical reactions quickly increase the mobile ion density when the cell is put into the reverse bias. The increased mobile ion density reduces the tunnel barrier for hole injection to less than a few nanometers and allows current to flow easily throughout the device.[42] In contrast, if the HTL covers the ITO effectively, as 35 nm of PTAA does, then the electrochemistry is mostly shut off. In that case -35 V is needed to force 19 mA/cm$^2$ through the device under reverse bias. This yields an electric field of $\sim 6 \times 10^7$ V/m in the perovskite layer (**Supporting Information Note 1**). We speculate that the breakdown occurs preferentially in spots where the perovskite is thinner,[28] the ITO exhibits asperities,[42] or similar defects are present in the transport layers. Nevertheless, we note that this electric field is comparable with the avalanche breakdown fields for both Si or GaAs over doping ranges from $1 \times 10^{16}$ to $4 \times 10^{17}$ cm$^{-3}$.[43] Given its stochastic nature, we expect avalanche breakdown can lead to failures similar to those in **Figure 1c**. While real devices will invariably contain microscopic inhomogeneities, and asperity-free cells will survive higher reverse bias,[28,30] we propose that even nearly-planar cells will thus eventually fail under reverse-current conditions in architectures that allow fields to approach avalanche breakdown levels. We emphasize that this effect isn't necessarily a drawback as much as a design tradeoff: designing cells to block reverse current until a large reverse bias is reached would allow them to operate reliably with bypass diodes that limit the possible reverse bias across a cell in a module.

On the other hand, if the goal is to operate with fewer (or even zero) bypass diodes, then the cell should be designed to pass higher reverse current uniformly (to avoid hot spots) at the lowest bias possible (to minimize power dissipation). To explore the effects of passing reverse current at lower biases without immediate catastrophic damage, we next replaced the PTAA with the popular MeO-2PACz hole-transporting interface modifier (**Figure 2a**) that is associated with much lower $V_{rb}$ (< |-5| V, **Figure S1**).[16] We posit that MeO-2PACz-based devices should behave very differently because MeO-2PACz does not uniformly cover the underlying ITO contact.[16,32,44,45] Consequently, reduction reactions can readily occur at this electrode under reverse bias.[16] The electrochemistry allows the mobile ion density to increase rapidly,[42] which reduces the tunnel barrier width for holes. Current flow throughout the cell and allows large reverse current (*i.e.*, 19 mA/cm$^2$) to be achieved at a relatively low voltage that is not sufficient for avalanche breakdown. As a result, for MeO-2PACz cells that do not have obvious pinholes (**Figure S3** and **Supporting Information Note 2**), we see no burn marks and a softer, more gradual reverse bias evolution (**Figure S4**). The change from catastrophic breakdown in thick PTAA cells to gradual, recoverable breakdown in MeO-2PACz cells occurs, in part, because the cell passes current at much lower biases. While it would thus be more difficult to protect MeO-2PACz-based cells with bypass diodes, recoverable performance after passing $J_{mpp}$ is a critical step in achieving a cell that can survive shading in a module without bypass diode protection.

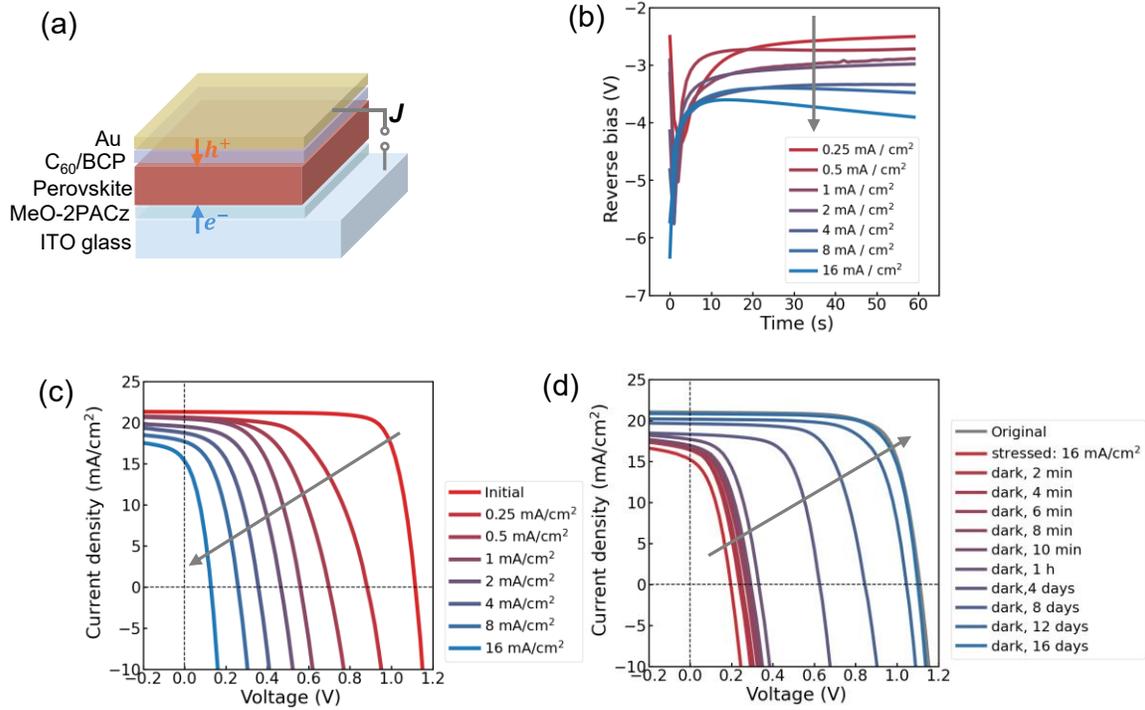

**Figure 2.** (a) Schematic diagram of a typical p-i-n structured MeO-2PACz-based perovskite solar cell soft-failures and recovery under current injection; (b) Reverse bias evolution of the perovskite solar cell under various fixed current conditions for 60 s; (c) $J – V$ curves of cells immediately after stressing at different reverse current density for 60s; (d) Recovery of the cells stored in the dark after stressing at 16 mA/cm² for 60s.

To better understand the degradation behavior of MeO-2PACz cells, we hold individual devices at progressively increasing current densities ($J$), ranging from 0.25 to 16 mA/cm², for 60 s in the dark. **Figure 2b** tracks the reverse bias required to drive the corresponding current flow. At all current densities, the bias first increases rapidly over the first ~1 s, reaches a maximum magnitude, then gradually falls to a stabilized value within around 10 s. We attribute this dynamic behavior during the first few seconds to a combination of ionic motion and redox processes like those described recently by Fremouw and co-workers, which shows mobile ion concentrations increasing by several-hundred times within 10 – 100 seconds under fixed reverse bias and moderate current densities (~100 µA/cm²).[42] Following their hypothesis, we propose three stages of early transient behavior corresponding to: (1) the first few hundred milliseconds, an interplay between the source-measure-unit setpoint feedback loop and drift of native mobile ions (halide vacancies) in the cell as the system reaches electrostatic equilibrium, forming of an ionic depletion layer at the ETL interface, (2) paired redox processes at both contacts leading to a rapid increase in the iodine vacancy concentration (falling bias from ~1 – 10 s), followed by (3) the establishment of a new steady state iodine vacancy concentration [36] resulting in a constant and relatively low voltage of -2 to -4 V.

Immediately after each stress condition, we measure $J – V$ curves under simulated 1-sun illumination to monitor the device performance changes (**Figure 2c**). With a constant stress duration of 60 s, the degradation increases with the magnitude of the applied $J$. Notably, among the various PV parameters (**Table S1**: including short-circuit current density ($J_{SC}$), open-circuit voltage ($V_{OC}$), fill factor (FF), fitted shunt resistance and series resistance (fitted in **Figure S5**)), we see most pronounced changes in $V_{OC}$, which we discuss in detail in a later section. Importantly, we find that all cells can recover their performance by resting in the dark,[29] with more seriously degraded cells taking longer to recover (**Figure 2d** and **Figure S6**). Illumination, forward bias, and thermal annealing all accelerate the cell recovery (**Figure S7**), unless the device's diode structure has been permanently damaged by catastrophic breakdown after hours of stressing, as can be reflected from the abrupt changes in reverse bias evolution (**Figure S8**).

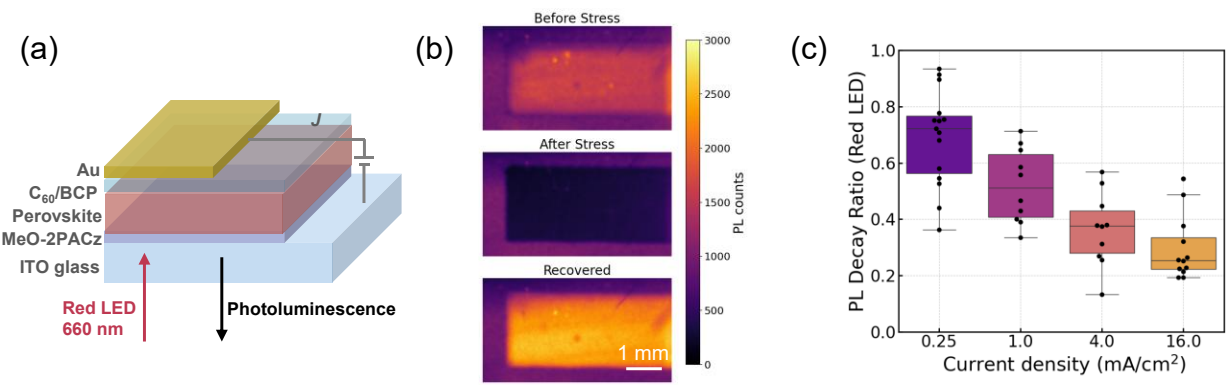

**Figure 3.** (a) Perovskite solar cells for widefield PL imaging; (b) The PL evolution of individual cells before and after sourcing reverse current ($J = 16$ mA/cm$^2$), under 660 nm red LED light illumination; (c) The PL decay ratios (quotients of post-stress and pre-stress PL intensities) as a function of applied current densities, under a fixed stressing time of 60 s.

To understand the device degradation behavior, we conducted widefield PL imaging on the perovskite solar cells before and after stress. Under red LED ($\lambda = 660$ nm) excitation (**Figure 3a**), we observe pronounced PL darkening across the whole device area after sourcing the cell with 16 mA/cm$^2$ reverse current for 60 s in the dark (**Figure 3b, top** and **middle**). Notably, the PL darkening is recoverable (**Figure 3b, bottom**), consistent with the recoverable device performance as discussed in **Figure 2d** and **Figure S7**. To quantitatively compare the PL darkening behaviors, we stressed 47 cells from 3 fabrication runs with 4 different current densities (16, 4, 1 and 0.25 mA/cm$^2$) for 60 s. We extract the average PL intensity of each cell and plot the PL decay ratios (defined as the quotients of post-stress and pre-stress PL intensities) against the applied current density. As shown in **Figure 3c** and **Table S2**, higher applied currents consistently lead to greater reduction in PL decay ratios.

Qualitatively, the increased PL quenching observed at higher reverse currents is consistent with the greater performance loss. Nevertheless, a closer comparison between the reduction in PL intensities (**Figure 3c**) and the decrease in $V_{OC}$s (**Figure 2c**) *shows that the extent of PL quenching is substantially smaller than what would be expected from the magnitude of $V_{OC}$ loss* (see **Supporting Information Note 3**, **Table S1** and **S2** for detailed analyses and information). This discrepancy suggests that, while reverse current stress obviously increases non-radiative recombination in the cells as observed by PL intensity, the $V_{OC}$ decrease, occurring at a faster rate than implied by the cells' radiative efficiency, must instead be governed by other processes. We speculate the ion motion and/or degradation of the interfaces may result in increasing deviation from ideal diode behavior, such that $V_{OC}$ is constrained well below the radiative limit in reverse-biased cells (see additional discussion in **Supporting Information Note 3**).

We also conducted additional control experiments to find out if the observed device performance change is due to low-resistance shunt formation[46] (*i.e.*, metal filaments[18,22]), rather than interfacial defect formation and electrochemical process. Importantly, the PL quenching (**Figure S9**) caused by the changing shunt resistances we extracted from the *J – V* curves (**Table S1** and **Supporting Information Note 4**) are not consistent with the measured PL quenching values (**Table S2**), indicating that the PL quenching is **not** primarily caused by changes in shunt resistance (see **Supporting Information Note 5** for detailed analyses). This result, which is thus supported by both the *J-V* curve fits of shunt resistance and the PL quenching ratios, further supports an electrochemical degradation mechanism, as distinct from the formation of simple shunt pathways.

Having examined the effects of varying reverse-current densities at fixed durations, we next investigate different combinations of current density and stress time such that the total injected charge (product of current density times duration) is equivalent. **Figure 4a** summarizes the device PCEs before (dot) and after (cross) reverse-current stress, as a function of cumulative injected charge density. The data largely clusters around a single trendline, suggesting that total reverse injected charge is the primary factor controlling device degradation. Closer inspection, however, reveals a finer and very informative structure within the data. **Figure. 4b** plots the normalized PCE (after-stress PCE divided by the initial PCE) for each device. After fitting with a single-exponential decay function (**Supporting Information Note 6**), the curves clearly separate out according to reverse current density. Notably, when compared on a per-injected-charge basis, the cells stressed with lower reverse bias currents show greater degradation. This trend is inconsistent with what one would expect from thermal damage and pure voltage- or field-driven degradation and argues strongly for electrochemically mediated degradation process under reverse-current injection. While more detailed analyses can be found in **Supporting Information Note 7**, here we highlight a few key takeaways. Under reverse bias, electronic charges (electrons or holes) can either be injected into the semiconductor's electronic bands or participate in electrochemical reduction or oxidation. We speculate that, as $J$ increases and $t$ decreases, the redox channel can become constrained by the local available ion concentration at the interface, which is limited by the ion's mobility, and the concentration gradient close to the perovskite/contact interface. In such a case, the redox process could become "transport limited": *i.e.* both migration and diffusion processes cannot keep up with the current demand.[33,34] Consequently, a smaller fraction of the injected charges participates in the redox processes, while a larger fraction is directly injected into the perovskite semiconductor bands, bypassing the degradation-inducing electrochemical pathways.

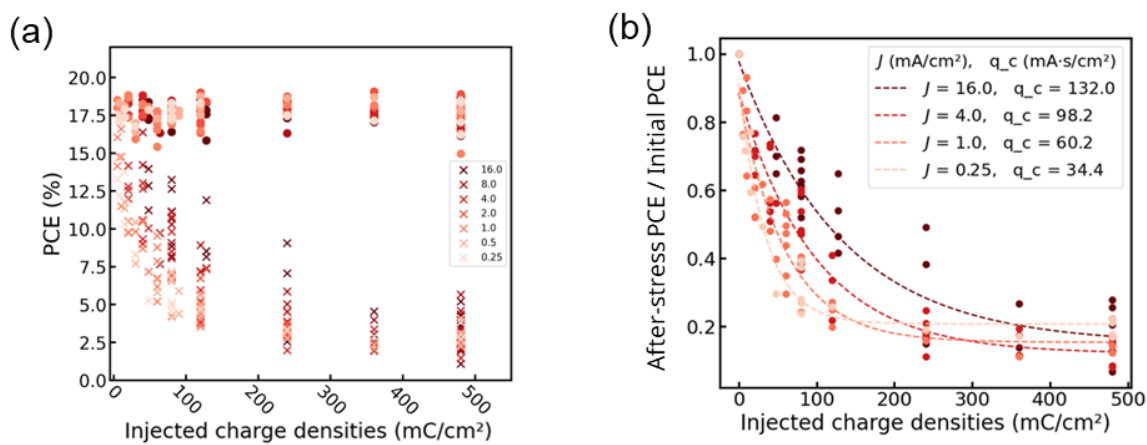

**Figure 4**. (a) PCEs statistics of over 250 individual cells from more than 6 fabrication runs before (dots) and after (cross) stress, and (b) degradation ratios and fitted single-exponential decay curves as a function of the total injected charge density ($J$ x $t$).

We note that all the above measurements and analyses apply to cells in complete darkness, generating zero photocurrent. However, this situation is not truly expected in real-world applications. For example, the shadow of a nearby tree will still allow some light to reach the cell from other angles, and the shadows often do not cover the entire area of just a single cell. In such cases, the shaded cell still generates photocurrent. If the resulting photogenerated carriers either worsen (or perhaps mitigate) the reverse current stability of the cell it could significantly complicate extrapolation of lab cells to real world conditions. Therefore, in **Supporting Information Note 11** and **Figure S14**, we systematically investigate cells under weak uniform illumination (**Figure S14a – 14c**) *or* with only a portion of their active area shaded (**Figure S14d – 14f**) under similar constant $J_{mpp}$ current conditions. In brief, this data set shows that the level of performance loss primarily increases with a decrease in the area-averaged illumination intensity over the entire cell area, whether it be from spatially non-uniform partial shading, or uniform decreased illumination over the entire cell area. In other words, these show that *device degradation correlates with the current deficit ($J_{mpp}$ - shaded cell photocurrent) of the low-photocurrent cells relative to their neighboring high-photocurrent cells, regardless of the shading conditions*. The photocurrent of the affected cells appears to be largely a parallel/independent process. This understanding should enable PV engineers to more reliably predict the degradation behavior of large perovskite solar cells operating in outdoor environments with diverse and complex shading conditions.

**Conclusion**

In this work, we systematically investigate the performance of p-i-n structured perovskite solar cells under reverse current injection conditions. We show that the choice of HTL dictates distinct device failure pathways: thick PTAA hole transport layers with better ITO planarization/electron blocking ability (thus high $V_{rb}$, >|-15| V) result in sudden catastrophic device failure, while MeO-2PACz phosphonic acid surface modifiers with poor ITO coverage (low $V_{rb}$, < |-5| V) lead to softer, more gradual device degradation that is mostly recoverable. For these cells, we find that PCE loss increases with the total injected charge deficit. For a given reverse injected charge level, we also find that the dynamics of cell degradation vary with the magnitude and duration of the reverse current, with smaller current injection for longer time periods resulting in more severe degradation than larger current injection for shorter time periods. This current dependence, together with the evidence of photoluminescence darkening in cells after charge injection, suggests that the degradation of perovskite solar cells by current injection is most likely redox reaction driven electrochemical processes. When large reverse currents are applied for short time, these electrochemical processes are likely constrained by transport and account for only a minor portion of the overall currents, leading to reduced degradation relative to smaller currents applied for extended durations. Ion mobility, electronic current, and electrochemical reactivity are thus tightly linked in perovskite cells under reverse bias and the observations presented here point toward potential pathways toward the characterization and improvement of perovskite active layer stability under reverse-current flow. We propose that efforts to demonstrate perovskite solar cells that survive or recover after reverse current stress should adopt strategies separate from those used to enable bypass diode usage. We anticipate such approaches are likely to involve: (1) selecting materials and device architectures to allow charge injection at low reverse bias, so as to avoid catastrophic device breakdown triggered by high fields across large charge injection barriers; (2) engineering the perovskite and interlayer properties to avoid undesirable redox reactions, such as tailoring defect densities and/or tuning the energies of charge injection so that any reverse injected charges are carried readily as electronic currents rather than driving redox reactions, and (3) ensuring that any products of the redox reaction be confined to the cell to allow their regeneration at a later time.

**Supporting Information**

Additional experimental procedures, materials, methods, additional data and analyses.

**Notes**

M.D.M. is an advisor to Swift Solar. The other authors declare no other competing interests.

**Author Contributions**

F.J. and D.S.G. conceived the project, designed the experiments and discussed the results together. F.J. and H.K. performed all the device fabrication, characterization and analyses. H.C. carried out the SEM measurement and assisted with the widefield PL measurements. R.G. helped with the oscilloscope measurements. Z.H. performed preliminary wide-field steady-state PL measurements. A.D. contributed to the $J – V$ curve fitting, wide-field PL measurement programming and interpretation. R.A.D., K.F., N.R.A., M.D.M. contributed to the data interpretation, manuscript review and editing. All authors contributed to the interpretation of the data as well as the presentation of this manuscript. All authors approved the submission. F.J. wrote the first draft. F.J and D.S.G. revised the manuscript with input from all the authors.

**Acknowledgments**

F.J. and H.C. acknowledge the support by the U.S. Department of Energy, Office of Basic Energy Sciences, Division of Materials Sciences and Engineering under Award DE-SC0013957 which primarily supported the work, including characterization and analysis of the samples. F.J. and D.S.G. additionally acknowledge support from the B. Seymour Rabinovitch Endowed Fund. H.K. acknowledges the Rabinovitch Endowed Fund for materials and supplies, as well as fellowship support from the Japan Patent Office for her role preparing samples; H.C. acknowledges a National Science Foundation Graduate Research Fellowship under Grant No. DGE-2140004. Any opinion, findings, and conclusions or recommendations expressed in this material are those of the author(s) and do not necessarily reflect the views of the National Science Foundation. The authors acknowledge the use of facilities and instruments at the Photonics Research Center (PRC) at the Department of Chemistry, University of Washington, as well as that at the Research Training Testbed (RTT), part of the Washington Clean Energy Testbeds system. F.J. acknowledges Dr. Farhad Akrami and Julisa Juarez for instrument assistance and fruitful discussions.